\documentclass[11pt]{article} 

\usepackage[utf8]{inputenc} 

\usepackage{geometry} 
\usepackage{graphicx} 

\usepackage{booktabs} 
\usepackage{array} 
\usepackage{paralist} 
\usepackage{verbatim} 
\usepackage{subfig} 
\usepackage{changes,cancel}
\usepackage{bbm}
\usepackage{amsmath,amsfonts,amssymb}
\usepackage{ulem}
\usepackage[active]{srcltx}
\usepackage{hyperref}
\hypersetup{colorlinks=true,pdfborder={0 0 0}}
\usepackage{fancyhdr} 
\usepackage{sectsty}
\allsectionsfont{\sffamily\mdseries\upshape} 

\usepackage[nottoc,notlof,notlot]{tocbibind} 
\usepackage[titles,subfigure]{tocloft} 

\def \be{\begin{equation}}
\def \ee{\end{equation}}
\def\id{\mathbbm{1}}

\newcommand{\bra}[1]{\left\langle #1 \right|}
\newcommand{\braket}[2]{\langle #1 | #2 \rangle}
\newcommand{\ket}[1]{\left| #1 \right\rangle}

\newcommand{\mbf}[1]{\mathbf{ #1} }
\newtheorem{con}{Conjecture}[section]
\numberwithin{equation}{section}

\begin{document}

\title{A study of the ambiguity in the solutions to the Diophantine equation for Chern numbers}

\author{J. E. Avron,  O. Kenneth and  G. Yehoshua }

\maketitle
\begin{abstract}
The Chern numbers for Hofstadter models with rational flux $2\pi p/q$ are partially determined by a Diophantine  equation. A $Mod~q$ ambiguity remains. The resolution of this ambiguity is only known for the rectangular  lattice with nearest neighbors hopping where it has the form of a ``window condition".
We study a Hofstadter butterfly on the triangular lattice for which the resolution of ambiguity is open. In the model many pairs $(p,q)$ satisfy a window condition which is shifted relative to the window of the square model.  However, we also find pairs $(p,q)$ where the Chern numbers do not belong to {\it  any contiguous window}.   This shows that  the  rectangular model  and the one we study on the triangular lattice are not adiabatically connected: Many gaps must close.  Our results suggest  the conjecture that the mod $q$ ambiguity in the Diophantine equation generically reduces to a sign ambiguity. 

\end{abstract}
\section{Motivation and results}

\begin{figure}[h!]
\begin{center}
\includegraphics[width=0.9\textwidth]{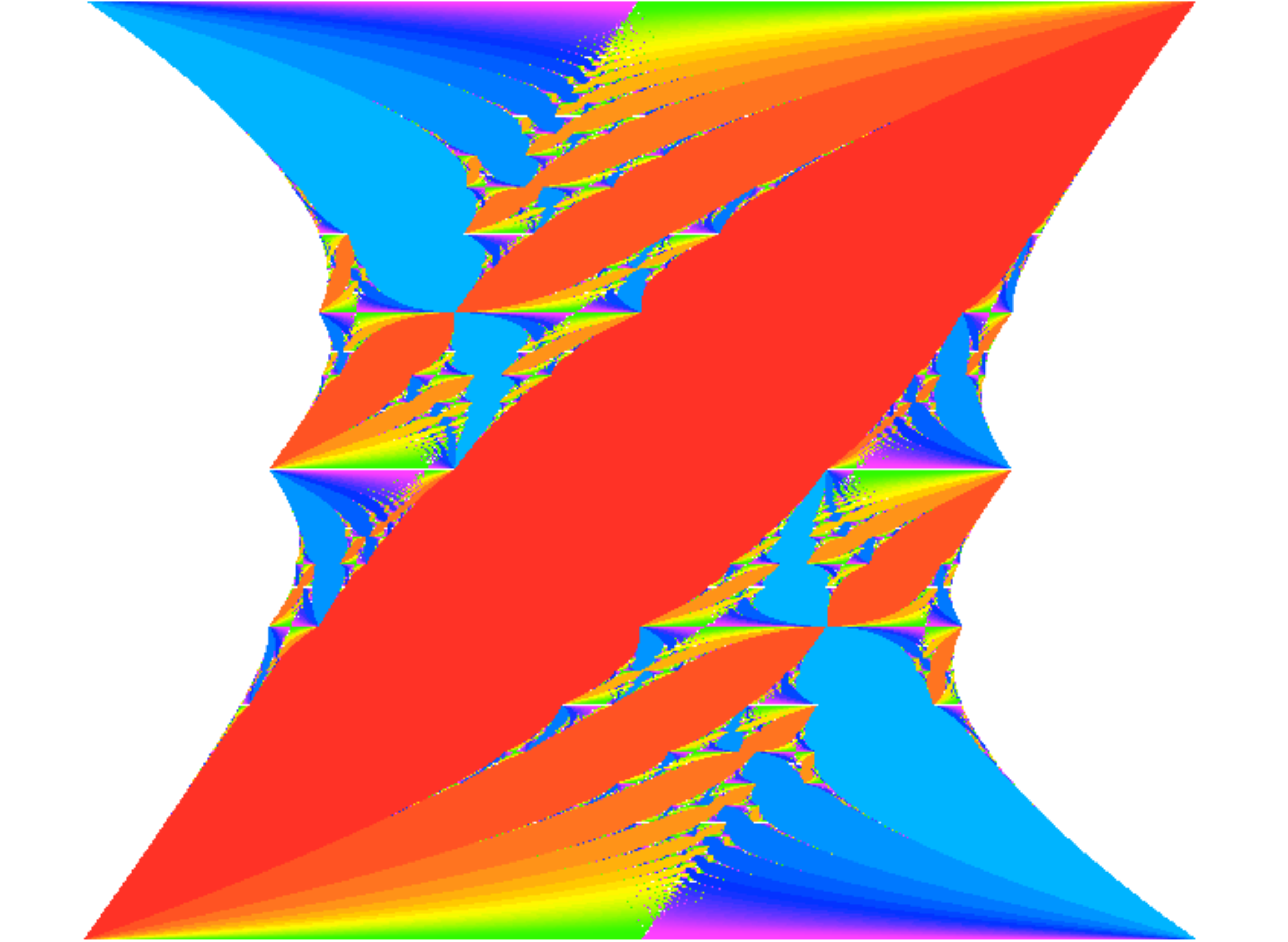}
\caption{A phase diagram for the Hofstadter model on  a triangular lattice where the flux through the down triangles $\Phi_d=\pi/2$.
The vertical axis is the total flux $\Phi$. The horizontal axis is the chemical potential. The colors represent the Chern numbers. The picture was made with the window condition Eq.\ref{t-window} for $q=512$ and $p$ ranging in [1,512].  The picture is apparently free from major coloring errors.  
}\label{Yeho}
\end{center}
\end{figure}

Hofstadter models give rise to  topological phase diagrams\footnote{The phase diagrams we consider should be distinguished from phase diagrams which describe the localization properties and the Liapunov exponent described e.g. in \cite{HanThouless}.}  with fractal structure \cite{avron2003topological,BellissardSimon}. The phases are labeled by the (integer) Hall conductances (Chern numbers).
A high resolution diagram, such as Fig.~\ref{Yeho},  requires efficient algorithms for approximating  the fractal spectrum of the Hofstadter models as well as an efficient algorithm to compute the  Chern numbers  that color  the wings of the butterfly. 

A numerical computation of the spectrum can be made efficiently for Hofstadter models that admit a ``Chambers relation" \cite{chambers}: A relation  that determines  the points in the (magnetic) Brillouin zone where gap edges occur. 
To color Fig. \ref{Yeho} which has, $q=O(550)$,  one needs   $O(10^5)$ Chern numbers. It is impractical to compute this many integers from their definition as integrals, Eq. (\ref{tknn}).  
One needs a shortcut. 
  
In the case of rational flux through the unit cell   
\be
\Phi=2\pi p/q, \quad  p,q\in \mathbb{N}, \ gcd(p,q)=1
\ee
  the Chern number of the j-th gap, $\sigma_j\in \mathbb{Z}$, satisfies the Diophantine equation:   
	\be\label{diophantine}
	\sigma_j= s\, j\mod~q,  
	\ee
$s$ is the modular inverse of $p$, i.e. $ sp=1\mod q$.  The gap indexes $j$ and $s$ take values in $\mathbb{Z}_q$ assuming that all the $q$  gaps  are open. 
The equation was first derived by a perturbation argument for the rectangular model in \cite{tknn}.  It was later shown to be a general result that holds for any periodic Schr\"odinger equation \cite{DAZ}.  In   Appendix \ref{sec:chern} we give a proof for tight-binding models.

The Diophantine equation forces distinct gaps to have different Chern numbers but  leaves a Mod $q$ ambiguity in $\sigma_j$  for $j\neq 0$. For $j=0$ and $j=q$, corresponding to  the semi infinite gaps below and above the spectrum, there is, of course, no ambiguity: $\sigma_0=\sigma_q=0$:  a trivial insulator \cite{bernevig}. 

The $Mod~q$ ambiguity had been resolved for Hofstadter model on the rectangular lattice with nearest nearest neighbors hopping in \cite{tknn}.  They showed  that, subject to the assumption  that  no gap opens or closes as the ratio of the horizontal to vertical hopping amplitudes changes away from zero,  $\sigma$ lies in a window:
\be\label{window}
	\sigma \in
\begin{cases}
	\left[1-\frac q 2 ,\frac q 2-1\right]& $q even$;\\
	\left[-\frac{q-1} 2,\frac{q-1} 2\right]& $q odd$\\
\end{cases}
\ee
 When $q$ is odd the window assigns $q$ values to the Chern numbers but when $q$ is even it only assigns $q-1$ values. This is still ok since the middle gap at $q/2$  (zero energy) is  permanently closed in the rectangular model.  The assumption that  no gap closes upon the special deformation of the rectangular  model used in  \cite{tknn}  was subsequently proved in \cite{choi,mouche1989coexistence}. This may be phrased as the statement that the Hofstadter models on the square and rectangular lattices  are  adiabatically connected.  

For  models, on other lattices, such as the Hofstadter model on the triangular lattice \cite{bks}, or on the hexagonal lattice \cite{ClaroWannier,Hatsugai74}, and models with hopping beyond nearest neighbors, the Diophantine equation still holds, but the issue of the mod $q$ ambiguity is open.  In all these models the $Mod~q$ ambiguity is a {\it finite ambiguity}  since the Chern numbers can be bounded in terms of the gap, see Eq.~(\ref{bd}) in Appendix \ref{cn}. However, 
the bound is not good enough to determine $\sigma$  uniquely.     A colored Hofstadter butterfly for the hexagonal model has been made in the diploma thesis of  Andrea Agazzi \cite{graf} where the Chern numbers were numerically computed using edge currents.  This approach is numerically intensive.

The triangular and hexagonal lattices can be viewed as deformations of the square lattice by  tuning the hopping amplitudes. For example, tuning the next nearest neighbor hopping amplitude along the north-west south-east bonds away from zero turns  the square lattice to  the triangular lattice. Two models are adiabatically connected if one can be deformed to the other without closing any gap where the resolution of the Mod $q$ ambiguity in the two models is the same.   However, there is normally no way of telling a-priori if all gaps remain open.  In fact, by the Wigner-von Neumann crossing rule, \cite{Friedland}, one would expect that generic deformations would  open and close some gaps\footnote{A generic deformation of Hofstadter models is associated with a   three parameters family:  Two parameters for the Bloch momenta and one for the deformation.  }. 

The Hofstadter model on the square lattice is not generic since its middle gap is closed  for all even $q$.  A generic Hofstadter model, though we can't put our hands on one, should have all its gaps open.   


One might think that one should be able to determine the Chern number easily from  Streda formula  \cite{streda} 
	\be\label{streda}
	2\pi  \delta\rho = \sigma\, \delta \Phi, \quad \rho=\frac  j q
	\ee 
Streda formula, however, comes with a catch: It  requires that one knows  {\it  a-priori} that two neighboring points $(\rho_1,\Phi_1)$ and $(\rho_2,\Phi_2)$ belong to the same wing of the butterfly.  Although humans can usually  correctly guess when two points belong to the same wing, it is an intuition that is difficult to translate to an algorithm that would allow a computer to make this guess. Once the resolution reduces to level of a single pixel, even humans can't guess. 
  
In this work we outline a graphic method of  to identify topological obstructions to adiabatic deformations which builds on the ability of humans to  solve  CAPTCHA (an acronym for {\it Completely Automated Public Turing test to tell Computers and Humans Apart}), which in this  case translates to  recognizing a coloring error.
 
We illustrate the method for the Hofstadter models on the triangular lattice \cite{HatsugaiKohmoto}.  In a triangular lattice there is a freedom to tune the fluxes in the up and down triangles, $(\Phi_u,\Phi_d)$.  This freedom allows for making various plots of infinitely many different  butterflies.  We have chosen to consider the case where the vertical axis in Fig. \ref{Yeho} is the total flux  $\Phi_u+\Phi_d$ and $\Phi_d=\pi/2$ is fixed.  We picked this particular value for $\Phi_d$ because it gives the butterfly  inversion symmetry. It  lacks the reflection symmetry   of  the rectangular and hexagonal lattices.

\begin{figure}[h!]
\begin{center}
\includegraphics[width=0.6\textwidth]{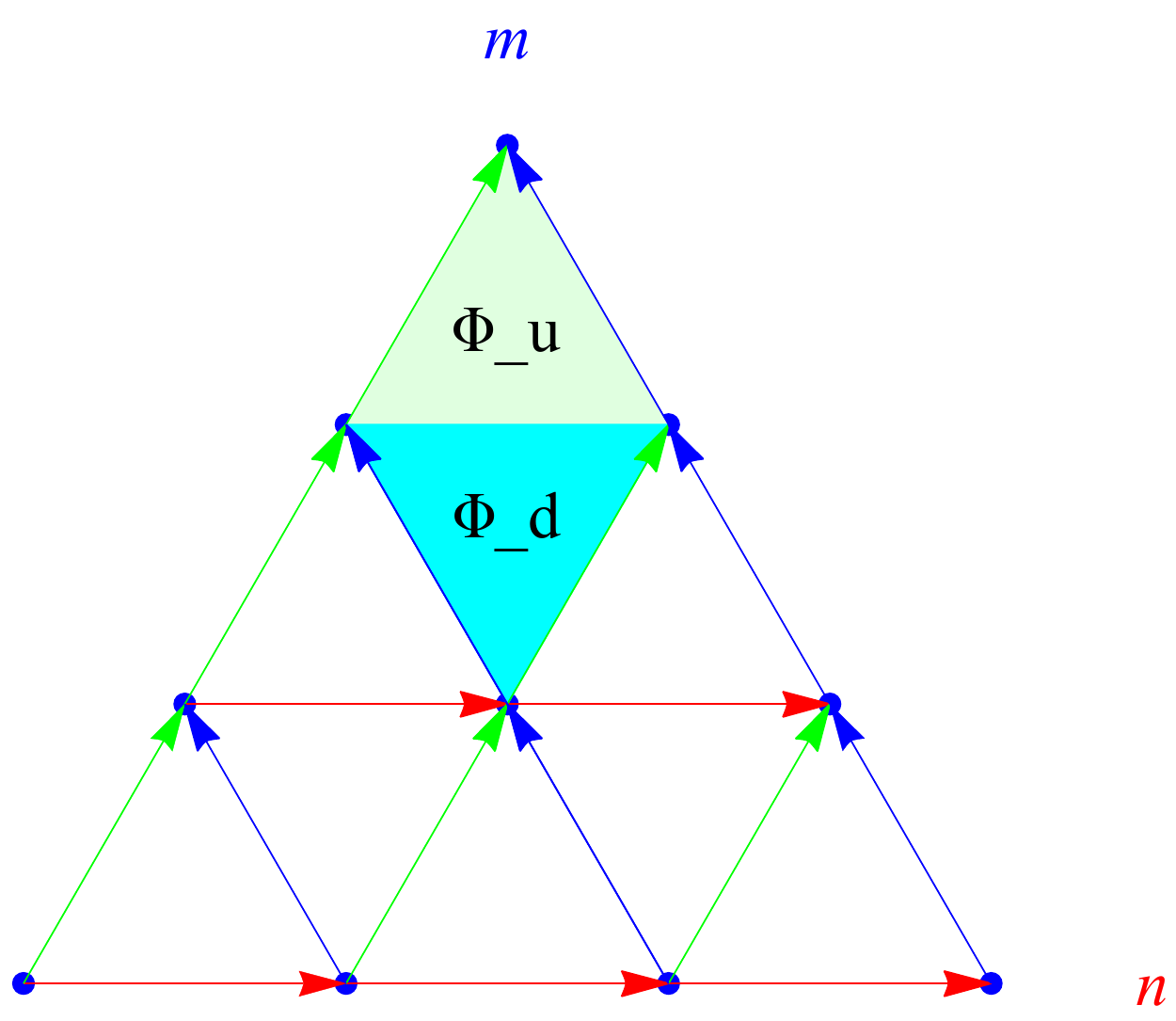}
\caption{Triangular lattice.
The flux through the down triangle  is   $e^{i\Phi_d}={\omega_d}$, and   the total flux through both up and down triangles is $\omega=e^{i\Phi}$. The coordinates $n$ grows towards the right and the coordinate $m$ grows towards the north-west.  $(n,m)$  is as in Eq. \ref{realize}.
}\label{lattice}
\end{center}
\end{figure}
 
 The Diophantine equation can be read as an assignment of a gap index $j$ to a given Chern number. The resolution of the ambiguity for a gap index is obvious, since $j\in 1,\dots, q$.  The ambiguity problem for $\sigma$ is now hidden in the fact that we do not know if a given $\sigma$, (rather than $\sigma\mod q$)  actually occurs.  
We know that $\sigma=0$ occurs. This suggests the heuristics that small Chern numbers $|\sigma|\ll q$ occur.  This is equivalent to saying that Eq. (\ref{window}) holds for $|\sigma|\ll q$ and fails for $|\sigma|=O(q)$.  An argument in favor of this heuristics can be made if one thinks of the Chern number as edge modes \cite{bernevig}. Generically, one expects edge modes to gap out so that their number is small.  

Asssuming this heuristics,     the Diophantine equation can be written graphically as
\be\label{chain}
\sigma_0=0 \rightarrow \sigma_p=1 \rightarrow  \dots \leftarrow\dots \sigma_{q-p}=-1\leftarrow\sigma_q=0
\ee
which resolves the ambiguity for small Chern numbers $|\sigma|\ll q$.  When the Chern number are $O(q)$ the assignment from the left and right in Eq. (\ref{chain}) disagree reflecting the mod $q$ ambiguity.

\begin{figure}[h!]
\begin{center}
\includegraphics[width=0.5\textwidth]{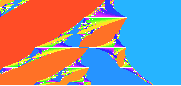}
\caption{The figure shows a  coloring mistake  for the pair $(p=2,q=5)$: A blue streak at the center of the figure, representing $\sigma=-2$, cuts the wing $\sigma=3$. The mistake reflects a wrong resolution of the Mod $q$ ambiguity of the solutions to the Diophantine equation.  
}\label{error}
\end{center}
\end{figure}

Fig. \ref{Yeho}, for the triangular lattice with $\Phi_d=\pm \pi/2$,  was plotted assuming the shifted window condition 
		\be\label{t-window}
	\sigma \in \left[-\frac q 2+1 ,\frac q 2\right], q ~even
\ee
The window appears to be free from {\it major} coloring errors. On the scale of few pixels, it becomes difficult to tell if the coloring is indeed right. The points $(q\pm 1)/ 2,q)$ were excluded because they lead to coloring errors  illustrated in Fig. \ref{error}. 

We have also   numerically computing the Chern numbers, Eq.~(\ref{tknn}) for few pairs $(p,q)$ with small $q$ and found:
\be\label{holes}
	\sigma \in
\begin{cases}
	[\cancel{-1}, 0,1], \ $gap ~closes$&     q=3,\ p\in\{1,2\} \\
	[\cancel{-2},-1, \dots ,3]&     q=5, \ p\in \{2,3\} \\
	\left\{- 4, \cancel{{-3}}, -2,\dots, 2, \cancel{{3}},4\right \}& q=7 $,\ $  p\in \{3,4\}  \\
	\{-4, \cancel{{-3}},-2,\dots 4,\cancel{{ 5}},6\}& q=9 $, \ $  p\in\{4,5\}\\
	\{-8,\cancel{{-7}},-6,\cancel{{-5}},-4,\dots, 4, \cancel{{5}}, 6,\cancel{{7}},8\} & q=13 $,\  $ p\in \{5,6\} \\
\end{cases}
\ee
This shows that there are pairs $(p,q)$ for which the Chern numbers {\it  do not} lie in any contiguous window.

  {\it There are topological obstructions} to  deformations Hofstadter models on the triangular lattice with $\Phi_d=\pm \pi/2$  to the square lattice: The two models are not adiabatically connected. This is true even if one restricts oneself to
 odd $q$ where all the gaps in the square model are open: The windows in Eq. (\ref{window}) is incompatible with Eq.~(\ref{t-window}) and Eq.~( \ref{holes}). 
 Most $(p,q)$  have gaps that  must  close. For example, 
the fragmented window $q=7$, results from a deformation of the contiguous window $[-3,3]$
upon gap closure taking $\pm 3 \mapsto \mp 4$. 


Our findings, Eqs.(\ref{t-window},\ref{holes}),  are consistent with the following conjecture:
\begin{con} The mod $q$ ambiguity in the solution of Eq.~(\ref{diophantine})  is, for generic Hofstadter models,  a sign ambiguity: The Chern number is either the smallest positive or the smallest negative solution of the Diophantine equation.  Equivalently: $-q\le \sigma\le q$.
\end{con}

The conjecture is related to interesting separate problem namely, how to determine the sign of Chern numbers. Determining the sign of an integral is, of course, a much easier problem than evaluating it and can be estimated, with high probability using Monte-Carlo methods.  In fact,  for small gaps, the sign of the Chern number is likely to be the sign of the curvature at the gap edged.  If the conjecture was true, it would allow for efficient algorithms for plotting high resolution Hofstadter butterflies when the resolution of the Mod $q$ ambiguity is not known.

In the appendixes we collect the tools we have used in the analysis. 


{\bf Acknowledgment} The research was supported by ISF. JA thanks Chris Marx, I. Dana, G. M. Graf and O. Zilberberg for  useful discussions.
\appendix

\section{Hofstadter models  on triangular lattice}\label{A}
	
Define  {\it magnetic hopping} $T_{1,2,3}$  on the triangular lattice, 
\be\label{realize}
(T_1\psi)(n,m)=\Psi(n,m-1), \ (T_2\Psi)(n,m)=\omega^m\Psi(n-1,m),\ T_3=\omega_u T_1 T_2
\ee
See Fig. \ref{lattice} for the meaning of $\omega,  \omega_d$, and the coordinates $(n,m)$.  
The unitary accumulated by going (clockwise) around the up/down triangles are $\omega_{u/d}$ and around the unit cell $\omega=\omega_u\omega_d$: 
\[
T_3^*T_2T_1=\omega_d,\quad T_3T_2^*T_1^*=\omega_u,\quad T_2T_1=\omega T_1 T_2 \quad 
\]
A (tight-binding) Hofstadter model  with isotropic hopping amplitudes is 
	\be\label{H}
	H(\omega,\omega_d)= T_1+T_2+T_3+h.c.
	\ee

\subsection{ $\Phi_d=\pi/2$: Inversion symmetry} \label{sec:inversion}
Hofstadter models on the triangular lattice give the freedom to choose independently the fluxes in the up and down triangles.  We have used this freedom to pick a model which is nice and symmetric.
    
The anti-unitary
\[ 
C\Psi(n,m)=(-)^{m+n}\bar \Psi(n,m)
\]
 acts on $H(\omega,\omega_d)$ by 
	\be\label{eh}
	 CH({\omega},{\omega_d})=-H(\bar {\omega},-\bar {\omega}_d)C 
	\ee
In a Hofstadter butterfly one looks at the spectrum as a function of the total flux $\Phi$.  It follows that $\omega_d=\pm i$ corresponds to a butterfly with  inversion symmetry of the two axes of the diagram: $(\Phi,E)\leftrightarrow (-\Phi,-E)$, a symmetry  evident in the Fig. \ref{Yeho}. 

\subsection{Reduction to one dimension}

The operators $T_j$ of Eq.~(\ref{realize}) are independent of the coordinate $n$.  The symmetry allows reducing the problem from two dimensions, $\mathbb{Z}^2$,  to one dimension, $\mathbb{Z}$. 
Let  $T$ and $S$ act  on the one dimensional lattice by 
\be\label{cr}
(T\psi)(m)=\psi(m-1),\quad (S\psi)(m)=\omega^m\psi(m),\quad ST=\omega TS
\ee
 Take $\Psi(n,m)= e^{-ik_1n} \psi (m)$  labeled by  
 the conserved (quasi) momentum $-\pi\le k_1\le \pi$. 
 One readily verifies that the action of $T_j$ on such functions takes the form
\begin{align*}
T_1\mapsto T,\quad T_2\mapsto e^{ik_1} S,\quad  T_3\mapsto \omega_ue^{ik_1} \,T S
\end{align*}
The Hofstadter Hamiltonian on $\mathbb{Z}^2$ has been reduced  to a periodic family of Hamiltonians, labeled by $k_1$, acting on $\mathbb{Z}$: 
\be\label{1d}
	H(k_1)= 
T (\id+ e^{ik_1} \omega_u S)+e^{ik_1}S+h.c., \quad |k_1|\le \pi
\ee
\subsection{Reduction to $q\times q$ matrices}
$T$ generates translations  and since it commutes with itself it is translation invariant.  $S$ is not.  However, when 
$\omega=e^{2\pi ip /q}$, a rational root of unity, $S^q=\id$. 
$H(k_1)$ is then periodic with period $q$.  This allows the reduction of the operator $H(k_1)$ acting on $\ell^2(\mathbb{Z})$  to a $q\times q$ matrix $H(k_1,k_2)$ parametrized by two quasi-momenta $\mbf{k}=(k_1,k_2) $.

  Let  $S$ and $T$ be  the $\mod~ q$ version of Eq.~(\ref{cr})
\be\label{ST}
S =\left(
 \begin{array}{cccccc}
{\omega}& 0&\dots&0&0\\
0&{\omega}^2&0&\dots&0\\
\dots&&\dots&\dots\\
0&0&0&{\omega}^{q-1}&  0\\
0&0&0&0&{\omega}^{q}
\end{array}
\right),
\
T=\left(
 \begin{array}{cccccc}
0& 0&0&\dots&1\\
1&0& 0&\dots&0\\
\dots&\dots&\dots&\dots&\dots\\
0&0&1&0& 0\\
0&0&0&1&0
\end{array}
\right)
\ee
The  $q\times q$ matrix obtained  from Eq.~(\ref{1d}) is
	\be\label{QP}
	H (\mbf{k})= e^{ik_2}T\left( \id + \omega_u  e^{ik_1}  S\right) +e^{ik_1} S+ h.c.
	\ee
The Bloch momenta $\mbf{k}$ takes values in the (Magnetic) Brillouin zone \cite{zak}
	\be
	BZ=\{\mbf{k}~|~|k_1|\le\pi, |qk_2|\le\pi\}
\ee
The matrices $S$ and $T$ satisfy the algebra
\be\label{Rot}
	ST={\omega}\ TS\, \quad S^q=T^q=\id
	\ee


\subsection{Magnetic symmetry}\label{sym}

The matrix $H (k_1,k_2)$ is {\it not} a periodic function on the (magnetic)  BZ. However, it is periodic up to a unitary transformation. In fact, there is a larger symmetry, known as ``magnetic symmetry" \cite{zak}. 

The commutation of $S$ and $T$, Eq.~(\ref{Rot}), imply
	\begin{align}\label{period}
	H (k_1,k_2)
=T^*H\left(k_1-\frac{2\pi p}q,k_2\right)T
= S^*H\left(k_1,k_2+\frac {2\pi p} q\right)\, S
	\end{align}
Since $gcd(p,q)=1$  $p$ has a modular inverse which we denote by  $s$.  Iterating  Eq. (\ref{period})  $s$ times give
\begin{align}\label{period-s}
	H (k_1,k_2)
=T^{s*}H\left(k_1-\frac{2\pi }q,k_2\right)T^s
= S^{s*}H\left(k_1,k_2+\frac {2\pi } q\right)\, S^s
	\end{align}
It follows that the spectral properties are fully determined  by a   small square in the BZ,  $\Omega_H$, whose size is  $2\pi/q\times2\pi/q$.
\section{Chambers relation and band edges}

An efficient computation of  the spectrum of Hofstadter models can be made provided there is a-priori knowledge where in the BZ band edges occur.   There is no known method to do that for general Hofstadter models, but Hofstadter models associated with tri-diagonal matrices are special. They admit Chambers relation \cite{chambers,HanThouless,bks} which facilitates this.  {Chambers } formula says that characteristic polynomial takes the form 
	\be\label{Chambers}
	\det \big(H(\mbf{k})-\lambda\big)= P(\lambda)+ \det H(\mbf{k})
	\ee
$P(\lambda) $ is a polynomial in $\lambda$ of degree $q$ which is  independent of $\mbf{k}$. 
This says that for all $p$ and $q$, band edges  occur at the extremal  points of $\det H(\mbf{k}).$ 

 For the triangular lattice with different fluxes in the up/down triangles \cite{bks} determined  $\det H(\mbf{k})$:
\begin{align}\label{chambers}
	\det H(\mbf{k})= h(\omega,\omega_d)+(-)^{q+1}\left(e^{i qk_1}+e^{i qk_2}+(-)^{q-1}\omega_u^q e^{iq(k_1+ k_2)}+ c.c\,\right)
	\end{align}

\subsection{Band edges for $\Phi_d=\pi/2$}

For $\omega_d=i$  the {extremal points of Eq. (\ref{chambers}) are determined by:
\begin{enumerate}
\item 
$q$ odd: The maximum and minimum of
\begin{align}\label{det}
	2\big(\cos x+\cos y\pm \sin( x+y)\big), \quad (x,y)= q\mbf{k}
	\end{align}
The band edges occur at
\be 
\pm q\mbf{k}\in (\pi/6,\pi/6), (5\pi/6,5\pi/6)
\ee 

\item  $q$ even: The maximum and minimum of
\begin{align}\label{det2}
	2\big(\cos x+\cos y\pm \cos( x+y)\big), \quad (x,y)= q\mbf{k}
	\end{align}
The band edges occur at 
\be  
\pm q\mbf{k}=(0,0), (2\pi/3,2\pi/3)
\ee 
\end{enumerate}

\section{Chern numbers}
 The adiabatic curvature of the n-th band is defined as
 \cite{berry}
 \be\label{omega}
  \Omega_n(\mbf{k})=2\ Im \ \braket{\partial_1\psi_n}{\partial_2\psi_n}= 2 \ Im\, \sum_{m\neq n} \frac{\bra{\psi_m} \partial_1H\ket{\psi_n}\bra{\psi_n} \partial_2H\ket{\psi_m} }{(E_n-E_m)^2}
 \ee
  The Chern number $\tilde\sigma_n$ associated with the $n$-th {\it band} is defined by \cite{tknn}
\be
\tilde \sigma_j=\frac 1 {2\pi} \int_{BZ}  \Omega_n(\mbf{k}) d^2k\in \mathbb{Z}
\ee 
The integration  is over the (magnetic) Brillouin zone. It is known to be an integer \cite{tknn}.  Using the magnetic symmetry, Section (\ref{sym}), it can be written as 
 \cite{tknn}
\be\label{tknn}
\tilde\sigma_j=\frac q {2\pi} \int_{BZ/q}  \Omega_n(\mbf{k}) d^2k=\frac q {2\pi i} \oint_{\partial(BZ/q)}\braket{\psi_j}{\nabla_k\psi_j} \cdot d\mbf{k}
\ee 
by Stokes formula.
\subsection{Chern numbers for gaps}\label{cn}
The Chern number $\sigma_j$ for the j-th gap is defined as the sum of Chern numbers of the bands below the gap: 
\be\label{bgap}
\sigma_j=\sum_{n\le j} \tilde\sigma_n
\ee 
The summand in Eq.~(\ref{omega}) is anti-symmetric under $m\leftrightarrow n$.  It follows that 
 \be\label{omega1}
\sum_{n\le j}  \Omega_n(\mbf{k})= 2 \ Im\, \sum_{n\le j<m} \frac{\bra{\psi_m} \partial_1H\ket{\psi_n}\bra{\psi_n} \partial_2H\ket{\psi_m} }{(E_n-E_m)^2}
 \ee

In the Hofstadter model, Eq.~(\ref{QP}),    $\partial_j H$ is a sum of six unitary operators and so   $\|\partial_j H\|\le 6$. It follows that with $g_j$ the gap
 \[
  \left|\sum_{n\le j}\Omega_n(\mbf{k})\right|\le  \frac {2\times 6^2} {g_j^2}  \sum_{n\le j<m}1= \frac{ 2\times 6^2 j(q-j) }{g_j^2}
 \]
 The area of BZ is $(2\pi)^2/q$ and hence
 \be\label{bd}
 |\sigma_j|\le \frac {4 \pi  \times 6^2 j(q-j)} {q \ g_j^2}
  \ee 

\section{ Diophantine equation}\label{sec:chern}

The n-th band  is associated with a  projection $P_n(\mbf{k})=\ket{\psi_n}\bra{\psi_n}$. We shall supress $n$ to simplify the notation.  $P$ is determined by the Hamiltonian and inherits its  symmetries. In particular,  $P(\mbf{k})$ is periodic, with period $2\pi/q$, up to the  unitaries $T^s$ and $S^s$ as per  Eqs.~(\ref{period-s}).  

The Chern number of the n-th band is, by Eq.~(\ref{tknn}), 
  $q/2\pi$ times the holonoly in the phase of  $\ket{\psi}$ as one parallel transports the state around the square 
\[(0,0)\to (2\pi/q,0)\to (2\pi/q, 2\pi/q)\to  (0,2\pi/q)\to (0,0) \]
 A parallel transport that keeps the Berry's phase, without accumulating a ``dynamical phase" is given by the  solution of the differential equation  \cite{kato50}
	\be\label{pt}
	\ket{d\psi}= i[dP,P] \ket{\psi}
	\ee
This evolution equation guarantees that $\ket{\psi}$ stays in  $Range~ P$, i.e. $P\ket{\psi}=\ket{\psi}$ all along the path: It satisfies the adiabatic theorem with no error.
  
Let $ \ket{\psi}_{(2\pi/q,0)}$ be the solution  of  Eq.~(\ref{pt})  along the open path $(0,0)\to (2\pi/q,0)$.  This  defines a phase $\gamma_1$ by
	\be\label{phase1}
 \ket{\psi}_{(2\pi/q,0)}= e^{i\gamma_1}T^s\ket{\psi}_{(0,0)}
	\ee
Similarly, let  $ \ket{\psi}_{(0,2\pi/q)}$ be the solution  of  Eq.~(\ref{pt})  along  the path $(0,0)\to (0,2\pi/q)$. It defines a phase $\gamma_2$ by 
		\be\label{phase2}
 \ket{\psi}_{(0,2\pi/q)}= e^{i\gamma_2}S^s\ket{\psi}_{(0,0)}
	\ee
Now, we can get two different determinations of $ \ket{\psi}_{(2\pi/q,2\pi/q)}$, one along the path $(0,0)\to (2\pi/q,0)\to (2\pi/q,2\pi/q)$ and the other along the path   $(0,0)\to (0,2\pi/q)\to (2\pi/q,2\pi/q)$. The discrepancy in the phases is the holonomy in phase associated with going around the square. This phase is precisely the value of the integral in Eq. (\ref{tknn}), which we are after. 

\begin{figure}[h!]
\begin{center}
\includegraphics[width=0.5\textwidth]{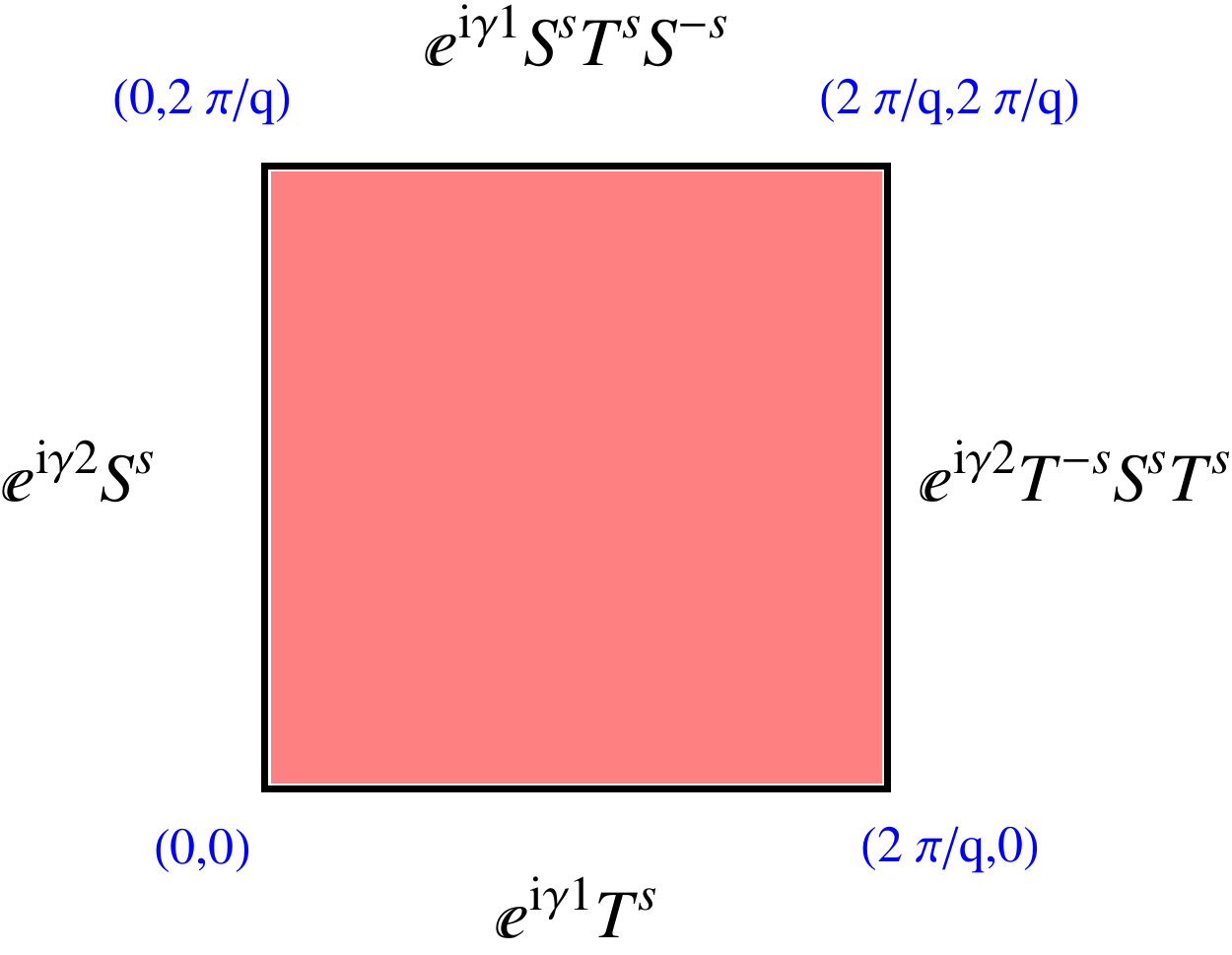}
\caption{The unitaries that relate $\ket{\psi}$ at the corners of the square $2\pi/q\times 2\pi/q$ by parallel transport along the corresponding edges.} \label{fig:dio}
\end{center}
\end{figure}

By the magnetic symmetry,  Eqs.~(\ref{period-s}),  parallel transport  along the path $(0,2\pi/q)\to (2\pi/q,2\pi/q)$ assigns the phase
		\be\label{phase3}
 \ket{\psi}_{(2\pi/q,2\pi/q)}= e^{i\gamma_1}S^sT^sS^{-s}\ket{\psi}_{(0,2\pi/q)}
	\ee
Similarly,  parallel transport  along the path $(2\pi/q,0)\to (2\pi/q,2\pi/q)$ assigns a different  phase to the same state
		\be\label{phase4}
 \ket{\tilde\psi}_{(2\pi/q,2\pi/q)}= e^{i\gamma_2}T^{-s}S^sT^s\ket{\psi}_{(2\pi/q,0)}
	\ee
Inserting Eq.~(\ref{phase1}) and  Eq.~(\ref{phase2}) in  Eq.~(\ref{phase3})  and  Eq. (\ref{phase4}) we find that the disagreement (holonomy) between these two assignments is 
	\be
 S^{-s}T^{-s}S^{s}T^s=(\omega^s)^s=e^{2\pi i s/q}
\ee
It follows from this and Eq.~(\ref{tknn}) that the Chern number of a single non degenerate band $j$ satisfies  the Diophantine equation: 	
	\[
	\sigma_j= s \mod q
	\]
This completes the proof of the Diophantine equation.



\end{document}